\documentclass[twocolumn]{svjour3}
\usepackage{graphics}
\usepackage{amsmath,amssymb}
\usepackage{epstopdf}
\usepackage{epsfig}
\journalname{Eur. Phys. J. C}

\begin{document}

\title{Possible $D^{(*)}\bar{D}^{(*)}$ and $B^{(*)}\bar{B}^{(*)}$ molecular states in the extended constituent quark models}

\author{You-Chang Yang\thanks{youcyang@163.com} \and Zhi-Yun Tan \and Jialun Ping\thanks{jlping@njnu.edu.cn (corresponding author)}
\and Hong-Shi Zong\thanks{zonghs@nju.edu.cn}}

\institute{You-Chang Yang \at Department of Physics, Nanjing University, Nanjing 210093,
School of Physics and Electrical Science, Zunyi Normal University, Zunyi 563006, and
State Key Laboratory of Theoretical Physics, Institute of Theoretical Physics, CAS, Beijing, 100190, China
\and Zhi-Yun Tan \at School of Physics and Electrical Science, Zunyi Normal University, Zunyi 563006, China
\and Jialun Ping \at Department of Physics, Nanjing Normal University, Nanjing 210023, P. R. China \and
Hong-Shi Zong \at Department of Physics, Nanjing University, Nanjing 210093, and
State Key Laboratory of Theoretical Physics, Institute of Theoretical Physics, CAS, Beijing, 100190, China}

\titlerunning{Possible $D\bar{D}$ and $B\bar{B}$ molecular states}
\authorrunning{Yang, Tan, Ping, Zong}

\maketitle

\begin{abstract}
The possible neutral $D^{(*)}\bar{D}^{(*)}$ and $B^{(*)}\bar{B}^{(*)}$ molecular states are studied
in the framework of the constituent quark models, which is extended by including the $s$-channel one
gluon exchange. Using different types of quark-quark potentials,
we solve the four-body Schr\"{o}dinger equation by means of the Gaussian expansion method.
The bound states of $D^{(*)}\bar{D}^{(*)}$ with $J^{PC}=1^{++},2^{++}$ and $B^{(*)}\bar{B}^{(*)}$ with
$J^{PC}=0^{++},1^{+-},1^{++},2^{++}$ are obtained. The molecular states $D^{*}\bar{D}$ with $J^{PC}=1^{++}$
and $B^{*}\bar{B}$ with $J^{PC}=1^{+-}$ are good candidates for the $X(3872)$ and $Z^0_b(10610)$, respectively.
\end{abstract}

\section{Introduction}
Since 2003, more than twenty new meson states (called $XYZ$ particles) \cite{chxPhysrept6391,ijmpa30,physrep6681} have been
observed by Belle, BaBar, BES, LHCb and other collaborations in hadronic final states that contain either
a $c\bar{c}$ or a $b\bar{b}$ quark pair. In general, the properties of these states do not match to the
expectations for any of the currently unassigned $c\bar{c}$ charmonium or $b\bar{b}$  bottomonium states.
A well established one among these $XYZ$ states is the $X(3872)$, which was first discovered in 2003 by
Belle Collaboration \cite{x3872prl91262001} in the $\pi^+\pi^-J/\psi$ invariant mass spectrum in
$B\rightarrow K\pi^+\pi^-J/\psi$, and later confirmed by six other experiments \cite{x3872prl93072001,x3872prl93162002,x3872epjc721972,x3872prd71071103,x3872jhep04154,x3872prl112092001}.
Its quantum number have been studied by Belle, BaBar, CDF and LHCb, and determined to be
$I^GJ^{PC}=0^{+}1^{++}$ \cite{x3872prl110222001}. The most striking feature of the $X(3872$ is the narrow
total width about $1.2$ MeV and the average mass $3871.69 \pm 0.17$ MeV, which is extremely close to the
$D^0\bar{D}^{0*}$ mass threshold \cite{PDG}.

Most of $XYZ$ states are unlikely interpreted as a conventional $c\bar{c}$ or $b\bar{b}$ meson for
their unusual properties. During past decades, various pictures like molecular state, compact tetraquark
state, hybrid state, and so on, have been proposed to explain the nature of them. For explaining the
structure of $X(3872)$, the most popular explanation is the molecular state. Swanson \cite{plb588189}
proposed to interpret the $X(3872)$ as a $D^0\bar{D}^{0*}$ molecular state with $J^{PC}=1^{++}$
which bound by both the pion and quark exchange. However, no $D^0\bar{D}^{0*}$ molecular state was obtained
in Ref.\cite{plb588189} if taking into account only of one pion exchange between $D^0$ and $\bar{D}^{0*}$.
Wong \cite{prc69055202} applied a quark-based model, which is similar to add short-range quark-gluon force,
to study the molecular sates composed of two heavy mesons. They found an S-wave $\bar{D}^{0}\bar{D}^{0*}$
molecular state with binding energy about 7.5 MeV. Suzuki \cite{prd72114013} believes that one pion exchange
potential can not bind $\bar{D}^0$ and $D^{0*}$ to molecular state. Thomas and Close \cite{prd78034007} found
that the $D^{0}\bar{D}^{0*}$ can be a bound state, when the pion exchange between charm and bottom mesons
is considered. However, their results are very sensitive to a poorly constrained parameter.
In Ref.\cite{epjc61411}, the author also obtained $D^{0}\bar{D}^{0*}$ bound state when they systematically
studied possible $D\bar{D}$, $D\bar{D}^{*}$ and $D^*\bar{D}^{*}$ molecular states by considered the vector,
pseudoscalar and scalar meson exchanges. In the framework of a potential model generated by the exchange
of scalar, pseudoscalar and vector mesons, which based on the effective Lagrangian of heavy hadron chiral
perturbation theory, a $D^*\bar{D}^{*}$ bound state was got by Lee, Faessler {\em et al}.\cite{prd80094005}
as well. In Ref.\cite{prd81014029}, the authors believe the $X(3872)$ should be understood as a molecular state 
of $D\bar{D}^{*}$, and extrapolates this information to make predictions of $B\bar{B}^{*}$ molecules 
\cite{prd88034018}. Gamermann, Oset {\em et al}.\cite{prd76074016} obtained a $D\bar{D}$ bound state both by 
a model using a chiral Lagrangian already used to study flavor symmetry breaking in Skyrme models, and another 
model by take into account a SU(4) symmetric Lagrangian with heavy meson-exchanges. They also analyzed the 
$e^{+}e^{-}\rightarrow J/\psi D\bar{D}, J/\psi D\bar{D}^{*}$ reactions of Belle, and found a hidden charm scalar 
meson with mass around 3700 MeV \cite{epja36189}, which is compatible with the $D\bar{D}$ bound state. 
In Ref. \cite{prd80114013}, Molina and Oset interpreted the $Y(3940), ~Z(3940)$ as molecular states of 
$D^*\bar{D}^{*}$ with quantum number $J^{PC}=0^{++},~2^{++}$ and $ X(4160)$ as a $D_s^*\bar{D}_s^{*}$ 
molecular state with $J^{PC}=~2^{++}$, respectively.

In constituent quark model, Vijande {\em et al.}\cite{PRD76094022} studied the four-quark system
$c\bar{c}n\bar{n}$ by means of the hyperspherical harmonic formalism. However no bound states have been found
whether taking into account the exchange of scalar and pseudoscalar mesons or not. Yang and Ping
\cite{ijmpcs291460227} systematically studied the $D\bar{D}$, $D\bar{D}^{*}$ and $D^*\bar{D}^{*}$ by means of
the Gaussian expansion method (GEM). No neutral bound state of $D^{(*)}\bar{D}^{(*)}$  was found as well.
Liu and Zhang \cite{PRC79035206} obtained a $D^{0}\bar{D}^{0*}$ bound state in a chiral quark model with
including $\pi, \sigma$, $\omega$ and $\rho$ meson exchanges in it.

In nature only the colorless hadron is allowed, so there is no one-gluon annihilation interaction between quark
and antiquark with the same flavor in a conventional colorless $q\bar{q}$ meson. However, the $s$-channel
one gluon exchange interaction can exist in the neutral $D^{(*)}\bar{D}^{(*)}$ and $B^{(*)}\bar{B}^{(*)}$ system
and maybe plays a important role for binding them, since the color structure of a four-quark state is much richer
than that of a $q\bar{q}$ conventual meson. Based on the Bhaduri, Cohler and Nogami model(BCN),
Wang {\em et al.} \cite{cpc34105} believe that the $s$-channel one gluon exchange interaction is important for binding a
$D^*\bar{D}^{*}$ molecular state, which is a good candidate for the $X(3872)$.

In this work, we would like to study the possible neutral molecular states $D^{(*)}\bar{D}^{(*)}$ and
$B^{(*)}\bar{B}^{(*)}$ by two constituent quark models, which are extended by including the one-gluon annihilation
interaction between $u\bar{u}$ or $d\bar{d}$ light quark pairs.  We solve the four-body Schro\"{o}dinger
equation by means of GEM, which is a high accuracy method for few-body systems developed by Kamimura, Hiyama
{\em et al.} \cite{gem} and extensively performed in studying the mass spectrum of multi-quark system \cite{prd80114023,prd81114025,jpg39105001,jpg39045001,prd86014008,prd88074007}.

This paper is organized as follows. After the introduction, we present the extended constituent quark models
in Sec.\ref{ECQM}. The wave functions of $D^{(*)}\bar{D}^{(*)}$ and $B^{(*)}\bar{B}^{(*)}$ are constructed by
considering the isospin, total angular momentum, color and the Gaussian expansion method and listed
in Sec. \ref{wf}. We summarize our numerical results and perform some analysis in Sec. \ref{result} and
draw some conclusions in Sec. \ref{summary}.

\section{The Constituent Quark Model with $s$-channel one Gluon Exchange\label{ECQM}}
\subsection{Bhaduri, Cohler and Nogami model}
This quark model was proposed by Bhaduri and collaborators \cite{Bhduri,Janc1}. The Hamiltonian takes the form,
\begin{equation}
H=\sum_{i=1}^4\left(m_i+\frac{\mathbf{p}_i^2}{2m_i}\right)-T_{c.m.}+\sum_{j>i=1}^4
(V_{ij}^C+V_{ij}^G)\label{bcnh}
\end{equation}
with
\begin{eqnarray}
V_{ij}^G & = & \alpha_s\frac{\boldsymbol{\lambda}_i^c\cdot\boldsymbol{\lambda}^c_j}{4}
\left(\frac{1}{r_{ij}}-\frac{1}{m_im_j}\frac{e^{-r_{ij}/r_0}}{r_0^2r_{ij}}
  \boldsymbol{\sigma}_i\cdot\boldsymbol{\sigma}_j\right),\\
V_{ij}^{C} & = & \boldsymbol{\lambda}^c_{i}\cdot \boldsymbol{\lambda}^c_{j}~(-a_{c}r_{ij}
  - \Delta), \label{bcnv}
\end{eqnarray}
where $r_{ij}=|\mathbf{r}_i-\mathbf{r}_j|$ and $T_{c.m.}$ is the kinetic energy of the center-of-mass
motion. $\boldsymbol{\sigma}$, $\boldsymbol{\lambda}$ are the SU(2) Pauli matrices and the SU(3)
Gell-Mann matrices, respectively. The $\boldsymbol{\lambda}~$ should be replaced by
$-\boldsymbol{\lambda}^{*}$ for the antiquark.

\subsection{The chiral constituent quark model(ChQM)}
The chiral constituent quark model(ChQM)~\cite{slamanca} includes Goldstone-boson exchange potential
in addition to color confinement potential and $t$-channel one-gluon-exchange (OGE) potential
between quarks (antiquarks). The chiral partner, $\sigma$-meson exchange potential,
is also introduced here, although its effect is still in controversy~\cite{sigma}. The Hamiltonian of
the ChQM used here is given as follows,
\begin{eqnarray}
H &=& \sum_{i=1}^4 \left( m_i+\frac{\mathbf{p}_i^2}{2m_i}
\right)-T_{c.m.} \nonumber \\
&+&\sum_{j>i=1}^4 (V_{ij}^G+V_{ij}^C+V_{ij}^\chi+V_{ij}^\sigma),
~~\chi=\pi,K,\eta, \label{h_cqm}
\end{eqnarray}
The OGE potential reads
\begin{eqnarray}
V_{ij}^G = \alpha_{s}
\frac{\boldsymbol{\lambda}^{c}_{i}\cdot\boldsymbol{\lambda}^{c}_{j}}{4}
\left[{\frac{1}{r_{ij}}}-{\frac{2\pi}{3m_im_j}}~(\boldsymbol{\sigma}_{i} \cdot
\boldsymbol{\sigma}_{j})~\delta(\mathbf{r}_{ij})  \right],  \label{gluon}
\end{eqnarray}
where, $T_{c.m.}$, $\boldsymbol{\sigma}$, $\boldsymbol{\lambda}$ have the same meaning as the above.
In non-relativistic quark model, the function $\delta(\mathbf{r}_{ij})$ should be regularized \cite{bhad,weinstein}.
It reads
\begin{equation}
\delta(\mathbf{r}_{ij})=\frac{1}{4\pi r_{ij}~r_0^2(\mu)}~
e^{-r_{ij}/r_0(\mu)},
\end{equation}
where $r_0(\mu)=\hat{r}_0/\mu$ and $\mu$ is the reduced mass of the interacting
quark/antiquark-quark/antiquark pair, $\hat{r}_0$ is a parameter to be determined from the experimental data.
In non-relativistic quark model, the wide energy covered from light to heavy quark requires an
effective scale-dependent strong coupling constant $\alpha_s$ in Eq.(\ref{gluon}) that cannot be
obtained from the usual one-loop expression of the running coupling constant because it diverges when
$Q\rightarrow\Lambda_{QCD}$. So one use an effective scale-dependent strong coupling constant given by
\begin{equation}
\alpha_s(\mu)=\frac{\alpha_0}{\ln\left[(\mu^{2}+\mu_0^2)/\Lambda_0^2\right]}~,
\end{equation}
where $\mu_0$ and $\Lambda_0$ are the parameters to be obtained by fitting the normal meson spectrum.

A screened potential simulating the results of unquenched lattice calculations is given by
\begin{equation}
V_{ij}^{C} =\boldsymbol{\lambda}^c_{i}\cdot \boldsymbol{\lambda}^c_{j}
~\{-a_{c}(1-e^{-\mu_c r_{ij}})+ \Delta\}, \label{confinement}
\end{equation}
where $\Delta$ is a global constant to be fixed from experimental data.

Due to the spontaneous breaking of original $SU(3)_L\otimes SU(3)_R$ chiral symmetry at some momentum scale,
the Goldstone meson exchange occurs between quarks (antiquarks). The potential takes the form
\begin{eqnarray}
& & V_{ij}^{\pi} =C(g_{ch},\Lambda_{\pi},m_{\pi}){\frac{m_{\pi}^{2}}{{12m_{i}m_{j}}}}
    H_1(m_{\pi},\Lambda_{\pi},r_{ij}) \nonumber \\
& & ~~~~\times (\boldsymbol{\sigma}_{i}\cdot \boldsymbol{\sigma}_{j})~\sum_{a=1}^3\lambda_{i}^{a}
    \lambda_{j}^{a}, \\ \label{pi}
& & V_{ij}^{\eta} =C(g_{ch},\Lambda_{\eta},m_{\eta}){\frac{m_{\eta}^{2}}{{12m_{i}m_{j}}}}
    H_1(m_{\eta},\Lambda_{\eta},r_{ij}) \nonumber \\
& & ~~~~\times (\boldsymbol{\sigma}_{i}\cdot \boldsymbol{\sigma}_{j})~\left[cos\theta_P(\lambda_{i}^{8}
    \lambda_{j}^{8})-sin\theta_P(\lambda_{i}^{0} \lambda_{j}^{0})\right], \\ \label{eta}
& & V_{ij}^{\sigma} = -C(g_{ch},\Lambda_{\sigma},m_{\sigma})~H_2(m_{\sigma},\Lambda_{\sigma},r_{ij}),
    \\ \label{sigma}
& & H_1(m,\Lambda,r)=\left[ Y(mr)-{\frac{\Lambda^{3}}{m^{3}}} Y(\Lambda r)\right], \\
& & H_2(m,\Lambda,r)=\left[ Y(mr)-{\frac{\Lambda}{m}} Y(\Lambda r)\right], \\
& & C(g_{ch},\Lambda,m)={\frac{g_{ch}^{2}}{{4\pi}}}{\frac{\Lambda^{2}}{{\Lambda^{2}-m^{2}}}} m,
\end{eqnarray}
where $Y(x)$ is the standard
Yukawa function defined by $Y(x)=e^{-x}/x$ and rest symbols have
their usual meaning. The chiral coupling constant $g_{ch}$ is
determined from the $\pi NN$ coupling constant through
\begin{equation}
\frac{g_{ch}^{2}}{4\pi }=\left( \frac{3}{5}\right) ^{2}{\frac{g_{\pi NN}^{2}%
}{{4\pi }}}{\frac{m_{u,d}^{2}}{m_{N}^{2}}},
\end{equation}
and flavor $SU(3)$ symmetry is assumed.


\subsection{$s$-channel one gluon exchange interaction}
\begin{figure}[htb]
 \begin{center}
  \resizebox{0.25\textwidth}{!}{\includegraphics{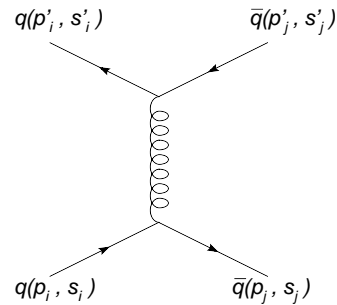}}
  \caption{The one-gluon annihilation diagrams for quark and antiquark.}
  \label{anni-gluon}
 \end{center}
\end{figure}

In the case of heavy-light meson and antimeson system, the contribution of $s$-channel annihilation
interaction should be taken into account. The one-gluon annihilation of light-quark and antiquark is
shown in Fig. \ref{anni-gluon}.
According to the Feynman rules, we can write down the $T$-matrix of the process
\begin{eqnarray}
T_{fi}  & = & \frac{g^2_s}{s} \bar{u} (p^{\prime}_i,s^{\prime}_i)
\chi^{\dag}_{c^{\prime}_i} \chi^{\dag}_{f^{\prime}_i}
\frac{\lambda^a}{2} \gamma^{\mu} v(p^{\prime}_j,s^{\prime}_j)
\chi_{c^{\prime}_j}\chi_{f^{\prime}_j} \nonumber \\
& & \bar{v}(p_j,s_j)\chi^{\dag}_{c_j} \chi^{\dag}_{f_j}
\frac{\lambda^{a}}{2}\gamma_{\mu}u(p_i,s_i)\chi_{c_i}\chi_{f_i},
\end{eqnarray}
where $s=(p_i+p_j)^2$ and $p$ is four-vector momenta; $u(p_i,s_i),~v(p_j,s_j)$ are the free Dirac spinors
of $i$th quark and $j$th antiquark; $\chi_{c},~\chi_{f}$ represent color and flavor wave function,
respectively. After Fierz transformation \cite{zong} of SU(n) group and taking the the lowest order
in the non-relativistic limit, the contributions from one-gluon annihilation to the potential
between quark and antiquark in momentum representation can be written as
\begin{eqnarray}
V_{ij}^{Anni-G}(s)&&=\frac{4 \pi
\alpha_s}{s}\frac{1}{4}\left(\frac{16}{9}-\frac{1}{3}\mathbf{\lambda}^{c}_{i}\cdot\mathbf{\lambda}^{*c}_{j}\right)
\nonumber\\
&&
\times\left(\frac{1}{3}+\frac{1}{2}\mathbf{f}^a_{i}\cdot\mathbf{f}^{*a}_{j}\right)
\left(\frac{3}{2}+\frac{\mathbf{\sigma}_{i}\cdot\mathbf{\sigma}_{j}}{2}\right).
\label{annieq1}
\end{eqnarray}
In coordinate space, and under the static approximation, $s=(m_i+m_j)^2=4m_q^2$ ($q=u$ or $d$ quark),
the potential reads
\begin{eqnarray}
&&V_{ij}^{Anni-G}(r_{ij})=\frac{\pi
\alpha_s}{4m_q^2}\left(\frac{16}{9}-\frac{1}{3}\mathbf{\lambda}^{c}_{i}\cdot\mathbf{\lambda}^{*c}_{j}\right)
\nonumber\\
&&~~~~~
\times\left(\frac{1}{3}+\frac{1}{2}\mathbf{f}^a_{i}\cdot\mathbf{f}^{*a}_{j}\right)
\left(\frac{3}{2}+\frac{\mathbf{\sigma}_{i}\cdot\mathbf{\sigma}_{j}}{2}\right)\delta(\mathbf{r}_{ij}).
\label{annieq2}
\end{eqnarray}
Here $\mathbf{f}^a$ is SU(3) matrix in the flavor space. The factor of first bracket represents that
this interaction never occurs inside color-singlet. Obviously, the last two factors in the brackets
mean that this interaction only occurs when the $\bar{q}q$ pair is in the same flavor with spin
$S = 1$. This interaction is always repulsive in molecular states of four-quark system,
since the color matrix elements is zero and $-\frac{14}{3}$ in $1\otimes1$ and $8\otimes8$,
respectively.

However, the earliest lattice simulations of gluon propagator in the Landau gauge, by Gupta {\em et al.}
\cite{PRD362813} were interpreted in terms of a massive particle propagator. In order to study the
$I=0~\pi\pi$ and $I=\frac{1}{2}~K\pi$ S-wave phase shift, Barnes and Swanson \cite{hepph9401326}
modified the gluon propagator by including an effective gluon mass. To analyze the mixing of the scalar
glueball with scalar-isoscalar quarkonia states above 1 GeV \cite{PRC71025202}, and investigate mesonic
content of the nucleon and Roper resonance \cite{PRC75067001}, the massive gluon propagator is also
employed. So here we choose the gluon propagator \cite{PRC71025202,PRep315,JHEP0401}
\begin{equation}
D(s)=\frac{1}{s-m_g^2},\label{propagator}
\end{equation}
where $m_g$ is effective gluon mass, which should be larger than the half of the bare glueball mass
deduced from lattice simulations. Typical values for the effective gluon mass are in the range
$0.6-1.2$~GeV  \cite{PRC71025202}.

After taking into account massive gluon propagator, the Eq.(\ref{propagator}) turns to be,
\begin{eqnarray}
&&V_{ij}^{Anni-G}(r_{ij})=\frac{\pi
\alpha_s}{4m_q^2-m_g^2}\left(\frac{16}{9}-\frac{1}{3}\mathbf{\lambda}^{c}_{i}\cdot\mathbf{\lambda}^{*c}_{j}\right)
\nonumber\\
&&~~~~~
\times\left(\frac{1}{3}+\frac{1}{2}\mathbf{f}^a_{i}\cdot\mathbf{f}^{*a}_{j}\right)
\left(\frac{3}{2}+\frac{\mathbf{\sigma}_{i}\cdot\mathbf{\sigma}_{j}}{2}\right)\delta(\mathbf{r}_{ij})
\label{annieq3}
\end{eqnarray}
Obviously, this interaction is attractive if $m_g>2m_q$.

\section{wave function\label{wf}}
The total wave function of four-quark system can be written as,
\begin{equation}
\Psi^{I,I_z}_{J,J_z}=\left|\xi\right\rangle
\left|\eta\right\rangle_{II_z}\Phi_{JJ_z},\label{twave}
\end{equation}
with\[\Phi_{JJ_z}=\left[\left|\chi\right\rangle_{S}\otimes\left|\Phi\right\rangle_{L_T}\right]_{JJ_z}
\]
where $\left|\xi\right\rangle$, $\left|\eta\right\rangle_{II_z}$, $\left|\chi\right\rangle_{SM_S}$,
$\left|\Phi\right\rangle_{L_{T}M_L}$ represent color, flavor, spin and spacial wave functions with quantum
numbers: color singlet, isospin $I$, spin $S$ and orbital angular momentum $L_T$, respectively.

The molecular states $D^{(*)}\bar{D}^{(*)}$ and $B^{(*)}\bar{B}^{(*)}$ system can be conveniently classified
in terms of total angular momentum, $J$, parity, $P$ and charge conjugation, $C$. In this work we only consider
the low-lying states, the orbital angular momentum $L_T$ is set to 0. In this case we have the following states
for the $D^{(*)}\bar{D}^{(*)}$ system:

(i) Two states with $J^{PC}=0^{++}$: $[D\bar{D}]_{0}$, $[D^*\bar{D}^*]_{0}$, where the subscript is total angular
momentum $J$.

(ii) One states with $J^{PC}=1^{++}$: $$\frac{1}{\sqrt{2}}\left([D\bar{D}^*]_{1}+[D^*\bar{D}]_1\right)$$
and two states with $J^{PC}=1^{+-}$: $$\frac{1}{\sqrt{2}}\left([D\bar{D}^*]_{1}-[D^*\bar{D}]_1\right)
 \mbox{ and } [D^*\bar{D}^*]_{1}$$.

(iii) One state with $J^{PC}=2^{++}$: $[D^*\bar{D}^*]_{2}$.

For the $B^{(*)}\bar{B}^{(*)}$ system, we replace the $D$ mesons in the above with the $B$ mesons.
The total spin function $\left|\chi\right\rangle_{SM_S}$ and flavor function $\left|\eta\right\rangle_{II_z}$
can be easily constructed from the above expressions. For example, for $[D^*\bar{D}^*]_{1}$,
\begin{eqnarray}
|\chi \rangle_{11} & = & \sqrt{\frac{1}{2}}(|11\rangle |10\rangle-|10\rangle |11\rangle) \nonumber \\
  & = & \frac{1}{2}(\alpha\alpha\alpha\beta+\alpha\alpha\beta\alpha
  -\alpha\beta\alpha\alpha-\beta\alpha\alpha\alpha) \nonumber \\
|\eta \rangle_{00} & = & \sqrt{\frac{1}{2}}(D^0\bar{D}^0+D^-D^+)
   =\sqrt{\frac{1}{2}}(u\bar{c}c\bar{u}+d\bar{c}c\bar{d}).  \nonumber
\end{eqnarray}

\begin{figure}[htb]
 \begin{center}
  \resizebox{0.25\textwidth}{!}{\includegraphics{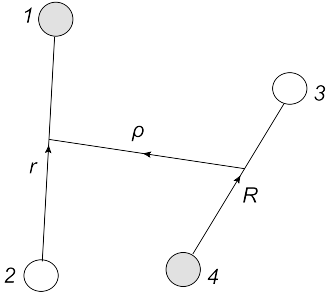}}
  \caption{The relative coordinates for the molecular state. Darkened and
   open circles represent quarks and antiquarks, respectively.}
  \label{jacobi}
 \end{center}
\end{figure}

The spatial structure of molecular states are pictured in Fig. \ref{jacobi}. We define the relative
coordinate as following,
\begin{eqnarray}
& & \mathbf{r}=\mathbf{r}_1-\mathbf{r}_2,\ ~~~~ \mathbf{R}=\mathbf{r}_3-\mathbf{r}_4,\\
& & \boldsymbol{\rho}=\frac{m_1\mathbf{r}_1+m_2\mathbf{r}_2}{m_1+m_2}
    -\frac{m_3\mathbf{r}_3+m_4\mathbf{r}_4}{m_3+m_4},
\end{eqnarray}
and the center of mass coordinate is
\begin{equation}
\mathbf{R}_{cm}=\sum_{i=1}^4 m_i \mathbf{r}_i/\sum_{i=1}^4 m_i,
\end{equation}
where $m_i$ is the mass of the $i$th quark(or antiqark).
Then the outer products of space and spin is
\begin{eqnarray}
\Phi_{JJ_z} & = & \left[ \left[ \left[ \phi^G_{lm}(\mathbf{r})\chi_{s_1m_{s_1}}\right]_{J_1M_1}
  \right. \right. \nonumber \\
 & & \left. \left. \left[ \psi^G_{LM}(\mathbf{R})\chi_{s_2m_{s_2}} \right]_{J_2M_2}
 \right]_{J_{12}M_{12}} \varphi^G_{\beta\gamma}(\boldsymbol{\rho})\right]_{JJ_z}.
\end{eqnarray}
Where $\chi_{sm_{s}}$ is spin wave function of normal meson which is composed by quark-antiquark.
The spatial wave functions $\phi^G_{lm}(\mathbf{r}), \psi^G_{LM}(\mathbf{R}),
\varphi^G_{\beta\gamma}(\boldsymbol{\rho})$ are written as,
\begin{eqnarray}
\phi^G_{lm}(\mathbf{r})=\sum_{n=1}^{n_{max}}c_nN_{nl}r^le^{-\nu_nr^2}Y_{lm}(\hat{\mathbf{r}})\label{gem1}
\end{eqnarray}
\begin{eqnarray}
\psi^G_{LM}(\mathbf{R})=\sum_{N=1}^{N_{max}}c_NN_{NL}R^Le^{-\zeta_NR^2}Y_{LM}(\hat{\mathbf{R}})\label{gem2}
\end{eqnarray}
\begin{eqnarray}
\varphi^G_{\beta\gamma}(\boldsymbol{\rho})=\sum_{\alpha=1}^{\alpha_{max}}c_{\alpha}
 N_{\alpha\beta}\rho^{\beta}e^{-\omega_\alpha \rho^2}Y_{\beta\gamma}(\hat{\boldsymbol{\rho}})\label{gem3}
\end{eqnarray}
Gaussian size parameters are taken as geometric progression
\begin{eqnarray}
\nu_{n}=\frac{1}{s^2_n}, & s_n=s_1 a^{n-1}, &
a=\left(\frac{s_{n_{max}}}{s_1}\right)^{\frac{1}{n_{max}-1}}
\label{geo progress}
\end{eqnarray}
The expression of $\zeta_N,\omega_\alpha$ in Eqs. (\ref{gem2})-(\ref{gem3}) are similar to
Eq. (\ref{geo progress}).

The color wave function of possible molecular states reads,
 \begin{eqnarray}
\left|\xi\right\rangle=&&\frac{1}{3}\left(\left|r\bar{r}r\bar{r}\right\rangle+
\left|g\bar{g}g\bar{g}\right\rangle+\left|b\bar{b}b\bar{b}\right\rangle
+\left|r\bar{r}g\bar{g}\right\rangle+\left|r\bar{r}b\bar{b}\right\rangle\right. \nonumber \\
&&+\left.\left|g\bar{g}r\bar{r}\right\rangle+\left|g\bar{g}b\bar{b}\right\rangle+
\left|b\bar{b}r\bar{r}\right\rangle+\left|b\bar{b}g\bar{g}\right\rangle\right),\label{color}
 \end{eqnarray}

\section{Numerical results and discussion\label{result}}
The energy of meson composed of quark-antiquark, and four-quark systems $D^{(*)}\bar{D}^{(*)}$ ,
$B^{(*)}\bar{B}^{(*)}$  can be obtained by solving the Schr\"{o}dinger equation
\begin{eqnarray}
\left(H-E\right)\Psi^{I,I_z}_{J,J_z}=0 \label{schrodinger}
\end{eqnarray}
with Rayleigh-Ritz variational principle.

To study the spectrum of a four-quark state, one believes that whether or not the state is bound
is judged by the threshold of two normal mesons, and the same parameters should used in the calculation
of normal mesons and four-quark states \cite{Janc1,weinstein,Manohar2,Bhduri2,Brink}.

For calculating spectrum of the normal meson, there is only one relative motion between quark and antiquark,
so the Eq.(\ref{gem1}) is employed. The model parameters of the ChQM and the BCN used in this work are shown in Table \ref{parameters}, which they are got from
Refs.\cite{Bhduri,Janc1,slamanca}, respectively. The calculated results of normal meson spectrum
listed in Table \ref{meson spectrum} are converged with $n_{max}=7$, $s_1=0.1$~fm and $s_{n_{max}}=2$~fm,
which are discussed in detail in Ref.\cite{prd80114023}. Obviously, the meson spectrum in Table
\ref{meson spectrum} calculated by GEM are agree well with the experimental data \cite{PDG} and Refs.\cite{Bhduri,Janc1,slamanca}.
\begin{table}[htb]
\begin{center}
\caption{Parameters of two quark models. ChQM: the masses of $\pi,\eta$ take the experimental
values, $m_\pi=0.7$~fm$^{-1}$, $m_{\eta}=2.77$~fm$^{-1}$; $m_{\sigma},\Lambda_{\pi}$,
$\Lambda_{\eta}$, $\theta_p$ are taken from Ref.\cite{slamanca}, namely $m_{\sigma}=3.42$~fm$^{-1}$,
$\Lambda_{\pi}=\Lambda_{\sigma}=4.2$~fm$^{-1}$, $\Lambda_{\eta}=5.2$~fm$^{-1}$, $\theta_p=-15^o$,
$g^2_{ch}/4\pi$=0.54. BCN: The parameters take from Refs.\cite{Bhduri,Janc1}}\label{parameters}
\begin{tabular}{clcc} \hline
Quark Model   &               & BCN  & ChQM \\ \hline
              & $m_{u,d}$     & 337   & 313    \\
Quark masses  & $m_s$         & 600   & 555    \\
   (MeV)      & $m_c$         & 1870  & 1752   \\
              & $m_b$         & 5259  & 5100   \\ \hline
              & $a_c$(MeV fm$^{-1}$) &176.738  &430    \\
  Confinement            & $\Delta$(MeV)      &-171.25  &181.1  \\
              & $\mu_c$(fm$^{-1}$)        &-    &0.7 \\ \hline
              & $\alpha_{s}$       &0.390209&        \\
              & $\alpha_{0}$       &-  &2.118   \\
     OGE      & $\hat{r}_{0}$(MeV fm)    & -   &28.17  \\
              & $r_{0}$(fm)    &0.4545   &  -  \\
              & $\mu_0$(MeV)       &-  &36.976 \\
              & $\Lambda_0$ (fm$^{-1}$)   & -  & 0.113 \\ \hline
\end{tabular}
\end{center}
\end{table}

\begin{table*}[htb]
\caption{Numerical results of normal meson spectrum (in MeV) for the ChQM and BCN models.
The column of BCN1 and ChQM1 are taken from Refs.\cite{Bhduri,Janc1} and \cite{slamanca}
respectively. The BCN2 and ChQM2 are calculated by GEM. The
last column takes from the latest Particle Data Group\cite{PDG}}\label{meson spectrum}
\begin{tabular*}{\textwidth}{@{\extracolsep{\fill}}cccccc@{}}
Meson          &BCN1 &BCN2 &ChQM1 &ChQM2 &Exp. \\ \hline
 $\pi$         &170 &137.5    &139    &153.2  &139.57$\pm$0.00035\\
 K             &537 &521.4    &496    &484.9  &493.677$\pm$0.016\\
 $\rho(770)$   &777 &779.6    &772    &773.1  &775.49$\pm$0.34\\
 $K^*(892)$    &905 &907      &910    &907.7  &896.00$\pm$0.25\\
 $\omega(782)$ &777 &779.6    &691    &696.5  &782.65$\pm$0.12\\
 $\phi(1020)$  &1018 &1018.5  &1020   &1011.9 &1019.422$\pm$0.02\\ \hline
 $\eta_c(1s)$  &3046 &3040    &2990   &2999.7 &2980.3$\pm$1.2\\
 $J/\psi(1s)$  &3102 &3098    &3097   &3096.7 &3096.916$\pm$0.011\\
 $D^0$         &1891 &1886.7  &1883   &1898.4 &1864.84$\pm$0.17\\
 $D^*$         &2021 &2021.3  &2010   &2017.3 &2006.97$\pm$0.19\\
 $D_s$         &2001 &1997    &1981   &1991.8 &1968.49$\pm$0.34\\
 $D^*_s$       &2103 &2102.3  &2112   &2115.7 &2112.3$\pm$0.5\\ \hline
 $B^\pm$       &5304 &5302    &5281   &5277.9 &5279.15$\pm$0.31\\
 $B^0$         &5304 &5302    &5281   &5277.9 &5279.53$\pm$0.33\\
 $B^*$         &5352 &5351.5  &5321   &5318.8 &5325.1$\pm$0.5\\
 $B^0_s$       &5376 &5373.1  &5355   &5355.8 &5366.3$\pm$0.6\\
 $B^*_s$       &5416 &5414.5  &5400   &5400.5 &5412.8$\pm$1.3\\
 $\eta_b(1s)$  &9431 &9422.2  &9454   &9467.9 &9399.0$\pm$2.3\\
 $\Upsilon(1s)$ &9448 &9439.5 &9505   &9504.7 &9460.30$\pm$0.26 \\ \hline
\end{tabular*}
\end{table*}

Generally, the binding energy of the four-quark system is defined by
\begin{equation}
\Delta E=E_T-E_{th}.
\end{equation}
with\[E_{th}=E_{M_1}+E_{M_2}\] where $E_{M}$ and $E_T$ represent the energy of
$Q\bar{q}~(Q=c,b$ and $q=u,d)$ and $Q\bar{q}q\bar{Q}$ systems, respectively. If $\Delta E < 0$,
then the system is stable against the strong interaction. According to the Table \ref{meson spectrum},
the thresholds of $S$-wave $D^{(*)}\bar{D}^{(*)}$ and $B^{(*)}\bar{B}^{(*)}$  of ChQM and BCN models
are listed in Table \ref{threshold}.
\begin{table}[htb]
\begin{center}
\caption{Threshold energies (in MeV) of $S$-wave $D^{(*)}\bar{D}^{(*)}$ and $B^{(*)}\bar{B}^{(*)}$ }. \label{threshold}.
\begin{tabular}{ccccc} \hline
 Configuration &$J^{PC}$&BCN &ChQM&Exp. \\ \hline
 $D\bar{D}$     &$0^{++}$          &3773.4   &3796.8  &3729.6\\
 $D^*\bar{D}$   &$1^{++},1^{+-}$   &3908.0   &3915.7  &3871.7\\
 $D^*\bar{D}^*$ &$2^{++}$          &4042.6   &4034.6  &4013.8\\ \hline
 $B\bar{B}$     &$0^{++}$          &10604.0  &10555.8 &10559.0\\
 $B^*\bar{B}$   &$1^{++},1^{+-}$   &10653.5  &10596.7 &10604.6\\
 $B^*\bar{B}^*$ &$2^{++}$          &10703.0  &10637.6 &10650.2 \\ \hline
\end{tabular}
\end{center}
\end{table}

To calculate the spectra of the four-quark states $D^{(*)}\bar{D}^{(*)}$ and $B^{(*)}\bar{B}^{(*)}$,
the Schr\"{o}dinger equation Eq.(\ref{schrodinger}) is solved by using the four-quark wavefunction
Eq.(\ref{twave}).
The converged results are obtained by taking the parameters of GEM as follows, $\alpha=12, ~n=7,~N=7$,
and the ranges of $s_n$ for $\boldsymbol{\rho}$ are from 0.1 to 6~fm, and 0.1 to 2~fm for
$\mathbf{R}$ and $\mathbf{r}$, respectively. Entem and Fern$\grave{a}$ndez believe the effective
gluon mass $m_g$ ranges from $0.6$~GeV to $1.2$~GeV \cite{PRC71025202}, so we calculated spectra of
$D^{(*)}\bar{D}^{(*)}$ and $B^{(*)}\bar{B}^{(*)}$ with $m_g$=0~GeV, 0.9~GeV, 0.97~MeV, 1~GeV
and without annihilation interaction. The results are listed in Table \ref{BCN_DDBB} and Table \ref{ChQM_DDBB}.
\begin{table}[hb]
\begin{center}
\caption{Energies (in MeV) of $S$-wave $D^{(*)}\bar{D}^{(*)}$ and $B^{(*)}\bar{B}^{(*)}$ with
different effective gluon mass for the BCN model. 'No-anni' means without annihilation interaction}
\label{BCN_DDBB}.
\begin{tabular}{ccccccc} \hline
Configuration  & $J^{PC}$   &$E_T$   & $\Delta E$ & $E_T$  & $\Delta E$  \\ \hline
  &     & \multicolumn{2}{c}{No-anni}      & \multicolumn{2}{c}{$m_g=0$GeV}  \\ \cline{3-6}
 $D\bar{D}$     &$0^{++}$   &3774.4   &1        &3774.6  &1.2    \\
 $D^*\bar{D}$   &$1^{++}$   &3909     &1        &3909.2  &1.2    \\
 $D^*\bar{D}$   &$1^{+-}$   &3909     &1        &3909.2  &1.2    \\
 $D^*\bar{D}^*$ &$2^{++}$   &4043.6   &1        &4043.7  &1.1    \\
 $B\bar{B}$     &$0^{++}$   &10604.3  &0.3      &10604.0 &0.4    \\
 $B^*\bar{B}$   &$1^{++}$   &10653.8  &0.3      &10653.9 &0.4    \\
 $B^*\bar{B}$   &$1^{+-}$   &10653.8  &0.3      &10653.9 &0.4    \\
 $B^*\bar{B}^*$ &$2^{++}$   &10703.3  &0.3      &10703.0 &0.5    \\ \hline
  &     & \multicolumn{2}{c}{$m_g=0.9$GeV} &\multicolumn{2}{c}{$m_g=1$GeV} \\ \cline{3-6}
 $D\bar{D}$     &$0^{++}$   &3765.1  &-8.3      &3773.1 &-0.3    \\
 $D^*\bar{D}$   &$1^{++}$   &3891.8  &-16.2     &3905.4 &-2.6    \\
 $D^*\bar{D}$   &$1^{+-}$   &3907.9  &-0.1      &3908.7 &0.7     \\
 $D^*\bar{D}^*$ &$2^{++}$   &4028.5  &-14.1     &4040.4 &-2.2    \\
 $B\bar{B}$     &$0^{++}$   &10659.3 &-34.7     &10591.4&-12.6   \\
 $B^*\bar{B}$   &$1^{++}$   &10608.9 &-44.6     &10634.5&-19 .0  \\
 $B^*\bar{B}$   &$1^{+-}$   &10643.4 &-10.1     &10650.9&-2.6    \\
 $B^*\bar{B}^*$ &$2^{++}$   &10650.5 &-42.5     &10684.9&-18.1   \\ \hline
\end{tabular}
\end{center}
\end{table}

\begin{table}[htb]
\begin{center}
\caption{Energies (in MeV) of $S$-wave $D^{(*)}\bar{D}^{(*)}$ and $B^{(*)}\bar{B}^{(*)}$ for
the ChQM. 'No-anni' means without annihilation interaction.} \label{ChQM_DDBB}.
\begin{tabular}{cccccc} \hline
Configuration  & $J^{PC}$   &$E_T$   & $\Delta E$ & $E_T$  & $\Delta E$  \\ \hline
  &     & \multicolumn{2}{c}{No-anni}      & \multicolumn{2}{c}{$m_g=0$GeV}  \\ \cline{3-6}
 $D\bar{D}$     &$0^{++}$    &3797.8    &1       &3798.0  &1.2   \\
 $D^*\bar{D}$   &$1^{++}$    &3916.6    &0.9     &3916.9  &1.2   \\
 $D^*\bar{D}$   &$1^{+-}$    &3916.8    &1.1     &3916.9  &1.2   \\
 $D^*\bar{D}^*$ &$2^{++}$    &4035.5    &0.9     &4035.8  &1.2   \\
 $B\bar{B}$     &$0^{++}$    &10554.6   &-1.2    &10556.2 &0.4   \\
 $B^*\bar{B}$   &$1^{++}$    &10592.4   &-4.3    &10597.2 &0.5   \\
 $B^*\bar{B}$   &$1^{+-}$    &10597.0   &0.3     &10597.1 &0.4   \\
 $B^*\bar{B}^*$ &$2^{++}$    &10633.8   &-3.8    &10638.1 &0.5   \\ \hline
  &     & \multicolumn{2}{c}{$m_g=0.9$GeV} &\multicolumn{2}{c}{$m_g=1$GeV} \\ \cline{3-6}
 $D\bar{D}$     &$0^{++}$    &3796.9  &0.1       &3797.3 &0.5       \\
 $D^*\bar{D}$   &$1^{++}$    &3913.2  &-2.5      &3915   &-0.7      \\
 $D^*\bar{D}$   &$1^{+-}$    &3916.7  &1.0       &3916.7 &1.0       \\
 $D^*\bar{D}^*$ &$2^{++}$    &4033.2  &-1.4      &4034.4 &-0.2      \\
 $B\bar{B}$     &$0^{++}$    &10537.3 &-18.5     &10543 &-12.8      \\
 $B^*\bar{B}$   &$1^{++}$    &10570.2 &-26.5     &10576.6 &-20.1    \\
 $B^*\bar{B}$   &$1^{+-}$    &10595.9 &-0.8      &10596.6 &-0.1     \\
 $B^*\bar{B}^*$ &$2^{++}$    &10613.4 &-24.2     &10619.4 &-18.2 \\ \hline
\end{tabular}
\end{center}
\end{table}

 From Table \ref{BCN_DDBB} and Table \ref{ChQM_DDBB}, we can find several intersting features.
 1) If we don't take into account the annihilation interaction, no bound stats of
 $D^{(*)}\bar{D}^{(*)}$ are found both in BCN and ChQM. However, there are three states
 of $B^{(*)}\bar{B}^{(*)}$ with $J^{PC}=0^{++},1^{++},2^{++}$ having the energies lower than
 the corresponding thresholds in ChQM, due to the larger mass of the $b$ quark than that of
 $c$ quark, which leads to the kinetic energy of former are lower than the latter. No state
 appears in BCN means that the meson-exchange interaction between $u(d,s)$ and $\bar{u}(\bar{d},
 \bar{s})$ plays an important role;
 2) If we don't take into account the effective mass of the gluon
 in annihilation interaction, namely $m_g=0$ in Eq.(\ref{annieq3}), no bound state of
 $D^{(*)}\bar{D}^{(*)}$ and $B^{(*)}\bar{B}^{(*)}$ are found both in BCN and ChQM since
 the annihilation interaction is repulsive in this case.
 3) With finite effective mass of gluon, almost all states under investigation can form
 molecules, having energies lower than the corresponding thresholds, and the binding
 energies decrease with the increasing effective mass of gluon. These results can be
 understood with the expression Eq.(\ref{annieq3}), where the magnitude of the denominator
 $4m_q^2-m_g^2$ increases with the increasing $m_g$ when $m_g> 2m_q \approx 626\sim 674$ MeV.
 So the results are sensitive to the effective mass of gluon. We choose $0.9\sim 1.0$ GeV for
 the effective gluon mass, which is in accord with the effective constituent gluon mass found
 in the study of gluon dynamics in Ref.\cite{PRD62074001} and the glueball-quarkonia content
 of scalar-isoscalar mesons in Ref.\cite{PRC71025202}. For the $D^{*}\bar{D}$ system,
 the $D^{*}\bar{D}$ with  $J^{PC}=1^{++}$  has about 2.5 MeV and 16.2 MeV binding energy for
 ChQM and BCN, respectively. The reason for the difference between two models is that the
 different masses of $u,d$ quark in BCN and ChQM are used, and the annihilation interaction
 Eq.(\ref{annieq3}) is sensitive to the quark mass. If we take 1 GeV for effective gluon mass
 in BCN which was chose in Ref. \cite{cpc34105}, the same binding energy as that of ChQM, 2.6 MeV
 is obtained. It is well known, the $X(3872)$ was first found by Belle Collaboration in the
 $J/\psi\pi^+\pi^-$ invariant mass spectrum in the decays of $B^\pm\longrightarrow K\pm J/\psi \pi^+\pi^-$.
 D0, BaBar, CDF, CMS, BESIII, and LHCb have later confirmed the $X(3872)$ by decays of $B^{\pm,0}$ masons
 and $pp$ collide, and affirmed the quantum number  $I^G(J^{PC})=0^+(1^{++})$. The average mass of $X(3872)$
 listed in PDG is $3871.69\pm0.17$ MeV, which is lower about 1 MeV than the threshold of $D^{*}\bar{D}$.
 Obversely, our result listed in Table \ref{BCN_DDBB} and \ref{ChQM_DDBB} is agree well with the
 experimental data when we choose reasonable parameters, which are suggested by Giacosa, Gutsche, and
 Faessler in Ref.\cite{PRC71025202}. The weakly bound state of $D^{*}\bar{D}^{*}$ with $J^{PC}=2^{++}$ are
 also obtained in our calculation. 4) For the $B^{(*)}\bar{B}^{(*)}$ system, four bound states are obtained
 when we take into account the annihilation interaction. The bound state of $B^{*}\bar{B}$ with
 $J^{PC}=1^{+-}$ is a good candidate for the $Z_b^0(10610)$, which is firstly found in the
 $\Upsilon(2,3S)\pi^0$ invariant mass spectrum in the
 $\Upsilon(10860)\rightarrow\Upsilon(1,2,3S)\pi^0\pi^0$ by Belle Collaboration\cite{PRD88052016}.

\section{Summary\label{summary}}
The constituent quark model is extended by introduced the $s$-channel one gluon exchange interaction,
which does not appear in the conventional mesons of $q\bar{q}$. We dynamically calculate the spectrum
of S-wave $D^{(*)}\bar{D}^{(*)}$ and $B^{(*)}\bar{B}^{(*)}$ system in the extended quark models.
The annihilation interaction is repulsive if we don't take into account the effective gluon mass in it.
However, if we take massive gluon propagator and reasonable effective gluon mass in the $s$-channel
one gluon exchange interaction, two molecular states $D^{*}\bar{D}$ with $I^G(J^{PC})=0^+(1^{++})$
and $B^{*}\bar{B}$ with $J^{PC}=1^{+-}$ are obtained, which they are good candidates for the $X(3872)$
and $Z^0_b(10610)$, respectively.  The $D^{*}\bar{D}^{*}$ , $B^{*}\bar{B}^{*}$ with $J^{PC}=2^{++}$,
$B^{*}\bar{B}$ with $J^{PC}=1^{++}$ and  $B\bar{B}$ with $J^{PC}=0^{++}$ are also predicted in these
extended constituent quark models. Further experimental searches by LHCb, BaBar, Belle and other
collaborations are needed to clarify whether these states exist or not.

In the present calculation, the one gluon annihilation interaction with effective gluon mass plays
an important role. The interaction does show up in the ordinary mesons even with the same flavor
because of the color structure. The effective mass of gluon is a model parameter, which does not
bound much by the experimental data. Because the calculated results are sensitive to the effective mass of
gluon, a better way to fix this parameter is expected.

\section*{Acknowledgment}
This work is supported partly by the National Science Foundation of China (under Contracts Nos.11265017,
11675080, 11535005, 11475085 and 11690030), and the China Postdoctoral Science Foundation (under Grant
No.2015M571727), and by the Guizhou province outstanding youth science and technology talent cultivation
object special funds (Grant No. QKHRZ(2013)28).


\end{document}